\documentclass[
reprint,
amsmath,amssymb,
amssthm, 
aps,
floatfix,
]{revtex4-2}

\usepackage{multirow}

\usepackage{soul,framed}
\usepackage[monochrome]{xcolor}
\usepackage{bbold}
\usepackage{stfloats}
\usepackage[normalem]{ulem}
\usepackage{ifthen}
\usepackage{enumerate}
\newboolean{underlined}
\usepackage{subcaption}
\usepackage{caption}
\setboolean{underlined}{false} 
\DeclareRobustCommand{\ul}[1]{\ifthenelse{\boolean{underlined}}{\uline{#1}}{#1}}

\newboolean{strikeout}
\setboolean{strikeout}{false} 
\DeclareRobustCommand{\st}[1]{\ifthenelse{\boolean{strikeout}}{
\ifmmode\text{\sout{\ensuremath{#1}}}\else\sout{#1}\fi
}{#1}}

\usepackage{amsmath}
\usepackage{braket}
\usepackage{physics}
\usepackage{natbib}
\usepackage{graphics}
\usepackage{graphicx}
\usepackage{appendix}

\usepackage{amsthm}
\usepackage{thmtools}

\declaretheoremstyle[
  headfont=\normalfont\bfseries,
  notefont=\normalfont\bfseries,
  bodyfont=\normalfont\itshape,
  headpunct=,
]{myplainTheo}

\theoremstyle{myplainTheo}

\usepackage{tikz,tikz-3dplot}
\usetikzlibrary{shapes,calc,positioning}
\usetikzlibrary{calc,intersections}
\usetikzlibrary {arrows.meta,bending}
\usetikzlibrary{external}

\begin{document}
\preprint{APS/123-QED}

\title{Geometrical derivation of Wigner's angle for arbitrary Lorentz transformations of massless particles}
\author{Isabella Cerutti}
\author{Petra F. Scudo}\email{petra.scudo@ec.europa.eu}

\affiliation{European Commission Joint Research Centre (JRC), 21027, Ispra, Italy}

\date{\today}

\begin{abstract}
This note summarizes the physics and mathematics of Lorentz transformations for massless particles, specifically for photons. We provide a complete analytical derivation of Wigner's little group matrix and a closed formula for the calculation of Wigner's angle for arbitrary Lorentz transformations. Our derivation highlights the geometrical content of the sequence of little group transformations leading to Wigner's matrix and links it to classical theorems in spherical trigonometry.   
\end{abstract}

\maketitle

In his seminal work~\cite{wigner1993unitary}, Wigner addressed the geometrical and kinematical principles of invariance arising from the laws of motion under the inhomogeneous Lorentz (or Poincaré) group. 
Nearly fifty years later, Berry discovered the existence of an additional phase in cyclic Hamiltonian quantum states, whose origin was geometrical or topological~\cite{berry1984quantal}. In this case, the phase depends on the path followed by a particle in parameter space for those parameters to which the Hamiltonian owes its time dependence. 

The Berry phase is commonly also referred to as the `\emph{Wigner phase}' in the literature. This interchange in terminology arises mainly from the connection with Wigner’s little group for massless particles. The link between the two phases was explicitly established by Lindner, Peres and Terno~\cite{lindner2003wigner} showing how the Berry phase may be derived as a Wigner angle, for massless particles, considering a closed curve in momentum space.

The Wigner rotation is the quantum effect that couples the transformation of a reference frame to the particle's spin, as defined in~\cite{noh2021quantum}). Indeed, it is related to the \textit{unitary irredudicible representations} of the inhomogenous Lorentz group in the Hilbert space of quantum particles (Wigner's classification is based on the value of the standard 4-momentum of the particle).    
The rotation can occur due to the change of reference axes from the observer to the particle (i.e., in the standard Lorentz transformation), as well as from the combination of two non-collinear Lorentz boosts. The rotation arising from the latter is known as  `\emph{Thomas rotation}' or `\emph{Thomas precession}'. The effect of Thomas precession on the Lorentz transformation of electromagnetic fields is investigated in~\cite{malhotra2020effect}. Closed-form expressions of the Thomas precession were derived using different approaches, based on hyperbolic geometry~\cite{vigoureux2001calculations,aravind1997wigner}, matrix formulations~\cite{kennedy2002thomasrotation}, \cite{caffarelli} and others like spinors and quaternions~\cite{van1984rotation,o2011elementary,berry2021lorentz}. 

The terms Thomas precession and Wigner angle have been often used interchangeably in literature, although their physical effects are distinct. Here we will refer to the product of successive boosts as Thomas precession and the result of the Lorentz transformation of a boosted particle as Wigner angle. As opposed to the Thomas precession~\cite{caffarelli}, the explicit calculation 
of the Wigner angle is more tedious due to the sequence of Lorentz transformations and therefore eluded by  researchers.
For massless particles, Caban and Rembieli{\'n}ski~\cite{caban2003photon} exploited the homomorphism between the group $SL(2,C)$ and the Lorentz group to explicitly derive the Wigner's little group elements. They conclude that the phase depends only on the direction of the momentum and does not depend on the frequency. This conclusion is also confirmed by Bergou, Gingrich and Adami~\cite{gingrich2003entangled}, who extended the analysis to the case of polarization-entangled photons.
Noh et al.~\cite{noh2021quantum} derived the Wigner angle induced by the gravitational field by assuming a Schwarzschild spacetime metric for a photon in proximity to Earth. In this work, we confine ourselves to special relativity.

\begin{figure}[htb]
\centering
    \subfloat[The shaded circle contains all the axis $\hat{n}_1$ of the family of rotations $R_{\hat{z} \xrightarrow{} \hat{p}} (\hat{n_1}, \phi_1)$ taking $\hat{z}$ to $\hat{p}$] {

\tdplotsetmaincoords{90}{90}
\resizebox{0.7\columnwidth}{!}{%
\begin{tikzpicture}[scale=2,tdplot_main_coords]
\def\myr{1}

\tdplotsetthetaplanecoords{90}
\tdplotdrawarc[tdplot_rotated_coords,thick,ball color=cyan, opacity=0.3, draw opacity=1, name path=black]{(0,0,0)}{\myr}{0}{360}{}{}

\tdplotsetrotatedcoords{75}{80}{0}
\tdplotdrawarc[tdplot_rotated_coords,ball color=blue, opacity=0.3, draw opacity=0.3]{(0,0,0)}{\myr}{0}{360}{below}{}
\tdplotdrawarc[tdplot_rotated_coords, color=blue, opacity=0.3]{(0,0,0)}{\myr}{0}{180}{above right}{}

%


 \tdplotsetrotatedcoords{180}{25}{0}
 \tdplotdrawarc[dashed,tdplot_rotated_coords,name path=red, color=gray, opacity=0.7]{(0,0,0)}{\myr}{0}{360}{above left}{} 
 \tdplotdrawarc[thin, tdplot_rotated_coords, color=gray]{(0,0,0)}{\myr}{-90}{90}{below left}{}


\draw[ thick, fill] (0,0,0) circle [radius={0.5pt}];
\draw[ thick, fill] (0,0,0.82) circle [radius={0.5pt}];

\draw[-{Stealth}, thick, fill] (0,0) --(0,-0.4,-0.38) node[pos=1, below] {$\hat{z}$};  
\draw[-{Stealth}, thick, fill] (0,0) --(0,0.94,-0.15) node[pos=1.15] {$\hat{p}$};    
\draw[-{Stealth}, thin, fill, blue] (0,0) --(0,0.3,-0.4) node[pos=0.93, below right] {$\hat{r}$};    
\draw[-{Stealth}, thin, fill, blue] (0,0) --(0,0,0.8) node[pos=1,  above  ] {$\hat{s}$};    
\draw[-{Stealth}, thick, fill, black] (0,0) --(0,0.24,0.2) node[pos=1.1,  right  ] {$\hat{n}_1$};    
\coordinate (O) at (0,0,0);
\node[black] at (0.0,0.24,-0.17)  {$\theta$};    
\node[black] at (0.0,0.7,-0.38)  {$\phi_1$};    

\tdplotsetrotatedcoords{45}{70}{0}
\tdplotdrawarc[->, thick, tdplot_rotated_coords, color=black, opacity=0.9, draw opacity=0.5]{(0,0,0)}{\myr*0.27}{35}{118}{below}

 \tdplotsetrotatedcoords{180}{25}{0}
\tdplotdrawarc[->, thin, tdplot_rotated_coords, color=black, opacity=0.9, draw opacity=0.7]{(0,0,0)}{\myr*0.99}{25}{-65}{below}

\end{tikzpicture}}\label{fig:rotFamily}}%
    \\
\subfloat[Illustration of the spherical and polar triangles]{



\tdplotsetmaincoords{90}{90}
\resizebox{0.7\columnwidth}{!}{%
\begin{tikzpicture}[scale=2,tdplot_main_coords]
\def\myr{1}

\tdplotsetthetaplanecoords{90}
%
\tdplotdrawarc[tdplot_rotated_coords,thick,ball color=cyan, opacity=0.3, draw opacity=1,]{(0,0,0)}{\myr}{0}{360}{}{}
%

\draw[top color=cyan, draw opacity=0.5,opacity=0.2, bottom color=blue, opacity=0.5]  (0,0, 0.8) to[bend right=10] (0,-0.5,-0.7) to[bend right=20] (0,+0.97,-0.16)   to[bend right =20] cycle;

\shade[top color=pink, draw opacity=0.4,opacity=0.8, bottom color=violet, opacity=0.6]  (0,0.25,-0.42) to[bend right=16] (0,0.86,-0.2) to[bend right=11] (0,0.40,0.44)   to[bend left=10.5] cycle;

\draw[-{Stealth}, line width = 0.6 pt, fill,red] (0,0) --(0,0.24,-0.41) node[pos=1.1] {};

\draw[-{Stealth}, line width = 0.6 pt, fill,red] (0,0) --(0,0.85,-0.225) node[pos=1.1]{};
\draw[-{Stealth}, line width = 0.6 pt, fill,red] (0,0) --(0,0.41,0.43) node[pos=1.18 ] {};

\draw[ -{Stealth}, thin, fill, blue] (0,0) --(0,0, 0.8) node[pos=1,black, right] {} ; 
\draw[-{Stealth}, thin, fill, blue] (0,0) --(0,-0.5,-0.7) node[pos=1.1,darkgray, right]{}; 
\draw[-{Stealth}, thin, fill,blue] (0,0) --(0,+0.96,-0.16) node[pos=1.1,darkgray] {};

\tdplotsetrotatedcoords{40}{51}{0}
\tdplotdrawarc[dashed,tdplot_rotated_coords,name path=blue, color=violet, opacity=0.5]{(0,0,0)}{\myr}{0}{360}{below}{}
\tdplotdrawarc[tdplot_rotated_coords, color=red]{(0,0,0)}{\myr}{40}{220}{above right}{}

\tdplotsetrotatedcoords{69}{-80}{0}
\tdplotdrawarc[thin, dashed,tdplot_rotated_coords,name path=white, color=violet,  opacity=0.5]{(0,0,0)}{\myr}{-10}{350}{below}{}
\tdplotdrawarc[thin, tdplot_rotated_coords, color=red]{(0,0,0)}{\myr}{-5}{175}{left}{}

\tdplotsetrotatedcoords{180}{25}{0}
\tdplotdrawarc[dashed,tdplot_rotated_coords,name path=red, color=violet, opacity=0.5]{(0,0,0)}{\myr}{0}{360}{above left}{} 
\tdplotdrawarc[tdplot_rotated_coords, color=red]{(0,0,0)}{\myr}{-90}{90}{below left}{}

\draw[ thick, fill] (0,0,0) circle [radius={0.5pt}];
\draw[ thick, fill] (0,0,0.82) circle [radius={0.5pt}];


\coordinate (O) at (0,0,0);
\node[rotate=23] at (0,0.56, -0.43) {$R_{\hat{z} \xrightarrow{}  \hat{p}}$};
\node[rotate=82] at (0,0.459, -0.05) {$R^{{-1}}_{{\hat{z} \xrightarrow{} \hat{\Lambda p}}}$};
 \node[rotate=-62] at (0,0.73, 0.2) {$\Omega$};

 \node[black] at (0,0.08,-0.25) {$\hat{z}$};  
 \node[black] at (0,0.22,-0.135) {$\hat{p}$};    
\node[black] at (0,0.21,0.25) {$  \hat{\Lambda p}$};    

\coordinate (O) at (0,0,0);
\node[black] at (0.0,0, 0.9) {$n_1$} ;  
\node[black] at (0.0,-0.54,-0.74)  {$n_2$};  
\node[black] at (0.0,+1.1,-0.2)  {$n_3$};    

\end{tikzpicture}}

\label{fig:sphere}}
\caption{Representation of the rotations on a unit sphere}
\end{figure}
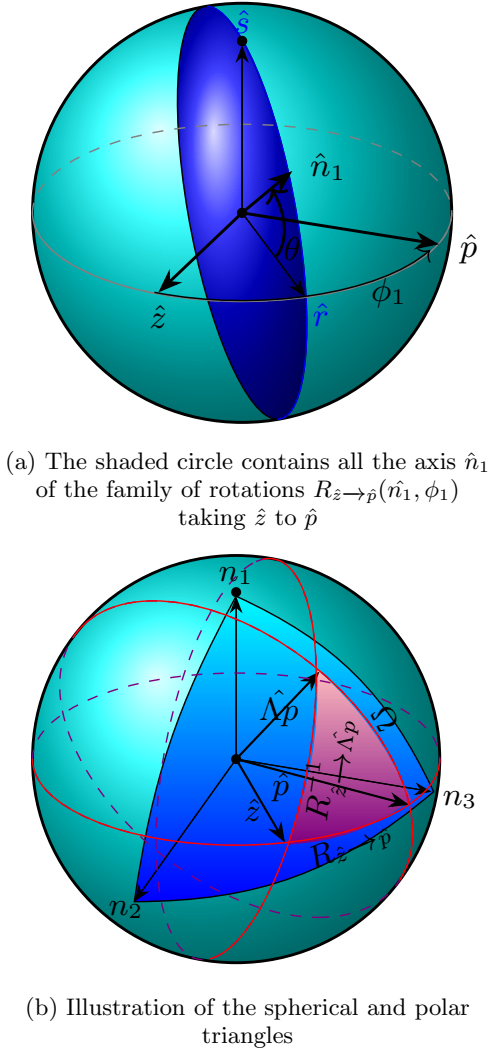

In this paper, the Lorentz transformation for massless particles is addressed starting from the seminal work of Weinberg~\cite{weinberg1995quantum} using a geometrical approach.
Given a reference system and a Thomas precession, we find that the Wigner angle is not unique, but varies as the rotation axis induced by the  standard Lorentz transformation (SLT) rotates completing a full circle in the plane orthogonal to the  axis involved in the SLT (Fig.~\ref{fig:rotFamily}).

In the case of SLT rotation with an axis orthogonal to the original and rotated vectors, an exact closed-form expression of the Wigner angle can be derived. 
The Wigner angle is  the  excess of the spherical triangle representing the motion of the particle in  momentum space starting and returning to the initial position.
The expression is presented here for the first time and confirms the earlier hypothesis by Lindner, Peres, and Terno\cite{lindner2003wigner} on the relation between topological phase and Wigner angle.


\section{Lorentz transformation}
The little group transformation, `Wigner matrix', is defined by Weinberg~\cite{weinberg1995quantum} as
\begin{equation}
    W(\Lambda,p) = L^{-1}(\Lambda p) \cdot \Lambda \cdot L(p) 
    \label{eq:LorentzTransf}
\end{equation}
where $L(p)$ is a standard Lorentz transformation carrying an initial 4-momentum $k=\omega_0 (1,0,0,1)$ along $\hat{z}$ - without loss of generality - onto a general 4-momentum
\begin{equation}
    p^\mu = p_0 (1,\hat{p})
\end{equation}
where $\hat{p}$ is given by
\begin{eqnarray}
    \hat{p} &=& (\sin(\xi) \cos(\psi), \sin(\xi) \sin(\psi), \cos(\xi)) \\
    L(p) &=& R(\hat{p}) B_z(p_0)
\end{eqnarray}
and $\Lambda $ is the generic Lorentz transformation.
Note that for the form of the standard Lorentz transformation, Weinberg assumes a rotation:
\begin{equation}
    R(\hat{p}) = e^{-i \psi J_3} ~  e^{-i \xi J_2},
\end{equation}

This way of expressing $R(\hat{p})$, originating from the angular momentum operators ($J_2$ and $J_3$), 
refers to the Euler angle representation. 

From now on, we will rather refer to the expression of rotation matrices as a function of a rotation axis ($\hat{n}$) and an angle ($\phi$).
Furthermore, Weinberg comments that the choice of the rotation $R(\hat{p})$ taking $\hat{z}$ to $\hat{p}$ is arbitrary and that any other choice of such a rotation taking $\hat{z}$ to $\hat{p}$ `would differ from the given one by at most a rotation around the $\hat{z}$ axis, corresponding to a mere redefinition of two phases of the one-particle states.'

We will delve into this point and analyse  the family of initial rotations transforming $\hat{z}$ into $\hat{p}$ and what the implications are of choosing a specific rotation within this family.

Upon rearranging the terms, Eq.~(\ref{eq:LorentzTransf}) becomes
\begin{equation}
    \Lambda \cdot L(p) = L(\Lambda p) \cdot W(\Lambda,p).
    \label{eq:Lp}
\end{equation}face
Let us assume that $\Lambda$ is a pure boost along a given direction $\hat{b}$, i.e., $\Lambda_{\hat{b}}$. To better understand the transformation, we will introduce explicit subscripts to indicate the starting and the final vector; we also omit the bold face notation for vectors. For instance, the rotation $R(\hat{p})$ carrying $\hat{z}$ onto $\hat{p}$  will be denoted as $R_{\hat{z} \xrightarrow{} \hat{p}}$. Hence, Eq.~(\ref{eq:Lp}) becomes
\begin{eqnarray}
     &&\Lambda_{\hat{b}}  \cdot   R_{\hat{z} \xrightarrow{} \hat{p}} \cdot  B_{\hat{z}}(p) \cdot R^{-1}_{\hat{z} \xrightarrow{} \hat{p}} \cdot R_{\hat{z} \xrightarrow{} \hat{p}} =   \label{eq:5} \\ 
     &&R_{\hat{z} \xrightarrow{} \hat{\Lambda p}} \cdot B_{\hat{z}}(\Lambda p) \cdot R^{-1}_{\hat{z} \xrightarrow{} \hat{\Lambda p}} \cdot R_{\hat{z} \xrightarrow{} \hat{\Lambda p}} \cdot W(\Lambda,p)  \nonumber
\end{eqnarray}
where we multiplied the right and left sides by a rotation matrix and its inverse, e.g., $R^{-1}_{\hat{z} \xrightarrow{} \hat{p}} \cdot R_{\hat{z} \xrightarrow{} \hat{p}} = \mathbb{1}$. The product $R_{\hat{z} \xrightarrow{} \hat{p}} \cdot B_{\hat{z}}(p) R^{-1}_{\hat{z} \xrightarrow{} \hat{p}}$  is a rotated boost along $\hat{p}$ direction and thus will be denoted as $B_{\hat{p}}$, leading to
\begin{equation}
     \Lambda_{\hat{b}}  \cdot  B_{\hat{p}}(p) \cdot R_{\hat{z} \xrightarrow{} \hat{p}} =  B_{\hat{\Lambda p}}(\Lambda p)  \cdot  R_{\hat{z} \xrightarrow{} \hat{\Lambda p}} \cdot W(\Lambda,p).
\end{equation}
By multiplying both sides by $R^{-1}_{\hat{z} \xrightarrow{} \hat{p}}$, we obtain
\begin{equation}
    \Lambda_{\hat{b}}  \cdot  B_{\hat{p}}(p) =  B_{\hat{\Lambda p}}(\Lambda p) \cdot R_{\hat{z} \xrightarrow{} \hat{\Lambda p}} \cdot  W(\Lambda,p) \cdot R^{-1}_{\hat{z} \xrightarrow{} \hat{p}} .
\end{equation}
The left side of this expression is the product of two non-commuting Lorentz boosts in directions $\hat{p}$ and $\hat{b}$. Thus, it follows the general decomposition of any Lorentz transformation~\cite{caffarelli} in the product of a boost times a rotation, or alternatively a rotation times a boost. This decomposition derives from the application of the polar decomposition theorem to the left hand side, leading to a unique rotation matrix $\Omega$ (see e.g.,~\cite{caffarelli,aravind1997wigner,moretti2006interplay})
\begin{equation}
    \Lambda_{\hat{b}} \cdot B_{\hat{p}}(p) = B_{\hat{\Lambda p}}(\Lambda p) \cdot \Omega
    \label{eq:polarDecomp}
\end{equation}
where $\Omega$ can be explicitly calculated as given in~\cite{caffarelli}.

Therefore, the Wigner matrix $W(\Lambda,p)$ is 
\begin{equation}
    W(\Lambda p) = R^{-1}_{\hat{z} \xrightarrow{}  \hat{\Lambda p}} \cdot \Omega \cdot R_{\hat{z} \xrightarrow{}  \hat{p}}.
\label{eq:W}
\end{equation}
From this equation, we see that the Wigner matrix is the product of three rotation matrices:
\begin{itemize}
    \item $R^{-1}_{\hat{z} \xrightarrow{} \hat{\Lambda  p}} (\hat{n}_3, \phi_3)$: the rotation carrying $ \hat{\Lambda p}$ to $\hat{z}$ with axis $\hat{n}_3$ and rotation angle $\phi_3$; 
    \item  $\Omega (\hat{n}_2, \phi_2)$: the rotation resulting from the non-commutativity of the two boosts with axis $\hat{n}_2$ and rotation angle $\phi_2$ as given in~\cite{caffarelli}. This rotation takes $\hat{p}$ to another vector $\hat{\Lambda p} =\Omega \cdot \hat{p}$. 
    \item  $R_{\hat{z} \xrightarrow{}  \hat{p}} (\hat{n}_1, \phi_1)$: the rotation carrying $\hat{z}$ to $\hat{p}$ with axis $\hat{n}_1$ and rotation angle $\phi_1$. 
\end{itemize}

The product of the three rotations brings the original vector $\hat{z}$ back to its initial position $\hat{z}$, indicating that $\hat{z}$ is an eigenvector of the product, i.e., the rotation axis of the product. 


\section{Computing the Wigner rotation}

The effect of the product of rotations in Eq.~(\ref{eq:W}) is topologically equivalent to moving on a sphere of unit radius (Fig.~\ref{fig:sphere}), bringing $\hat{z}$ to $\hat{p}$ with $R_{\hat{z} \xrightarrow{}  \hat{p}}$, then $\hat{p}$ to a new point ${  \hat{\Lambda p}}$ with $\Omega$ and finally to $\hat{z}$ again. The formed edges have a length equal to the cosine of the half angle between the vectors at the end-points. The angles are halved as the rotation product is an element of the SU(2) Lie group, double covering of $SO(3)$.  
The third rotation $R^{-1}_{ \hat{\Lambda p} \xrightarrow{}  \hat{z}}$ closes the path on the sphere, forming a spherical triangle~\cite{wick1962angular}. 
Therefore, the sequence of standard Lorentz transformations defining Wigner's little group is a realization of the closed path in the  momentum space of the system.

The axis and the angle of the Wigner rotation can be mathematically computed by resorting to quaternions $q_i= \cos(\phi_i/2) + n_i \sin(\phi_i/2), i=1, 2, 3$. The quaternion associated with the Wigner rotation, $q_W$, is thus defined by the product
\begin{equation}
    q_W= q_3 \cdot q_2 \cdot q_1.
\end{equation}
It is well known that for the product of two quaternions $q_c= q_2 \cdot q_1$, the following relationships hold
\begin{eqnarray}
  &\cos {\frac {\phi_c}{2}}=\cos {\frac {\phi_2 }{2}}\cos {\frac {\phi_1 }{2}}-\hat{n_2}\cdot \hat{n_1} \sin {\frac {\phi_2 }{2}}\sin {\frac {\phi_1 }{2}} \\
   &\hat{n_c}  \tan {\frac {\phi_c }{2}}={\frac {\hat{n_2} \tan {\frac {\phi_2 }{2}}+\hat{n_1} \tan {\frac {\phi_1 }{2}}+\hat{n_2} \times \hat{n_1} \tan {\frac {\phi_2 }{2}}\tan {\frac {\phi_1 }{2}}}{1-\hat {n_2} \cdot \hat{n_1} \tan {\frac {\phi_2 }{2}}\tan {\frac {\phi_1 }{2}}}}.
\end{eqnarray}
The relation of the last equation as application of quaternion products may be equally derived applying the \textit{spherical law of cosines} to the spherical triangle as defined by Rodrigues' vectors, starting from $\hat{z}$. The correspondence between quaternions and spherical triangulation is described in~\cite{McCarthy2005}. 
By extending the calculation to the product of three quaternions, it is possible to find the expression for the Wigner angle as:
\begin{eqnarray}
     \cos(\phi_W/2)&=& \cos(\phi_1/2) \cos(\phi_2/2) \cos(\phi_3/2) \label{eq:phi_W} \\
     &-& \sin(\phi_1/2) \sin(\phi_2/2) \cos(\phi_3/2) \hat{n}_1 \cdot \hat{n}_2 \nonumber \\
     &-& \sin(\phi_2/2) \sin(\phi_3/2) \cos(\phi_1/2) \hat{n}_2 \cdot \hat{n}_3\nonumber \\
     &-& \sin(\phi_3/2) \sin(\phi_1/2) \cos(\phi_2/2) \hat{n}_3 \cdot \hat{n}_1 \nonumber \\ 
     &- &\sin(\phi_1/2) \sin(\phi_2/2) \sin(\phi_3/2) \hat{n}_3 \cdot( \hat{n}_2 \cross \hat{n}_1) \nonumber
\end{eqnarray}
which is invariant to cyclic permutations and can be applied to Eq.~(\ref{eq:W}) in straightforward manner, after computing the axis $\hat{n_i}$ and angles of rotation $\hat{\phi_i}, i=1, 2, 3$.
We recall that the Wigner matrix $W(\Lambda p)$ represents an element of the little group of which the unitary transformation $U(W)$ in the particle's Hilbert space is a function. For a massless particle with momentum $p$ and helicity $\sigma$ in a state $\Psi_{\sigma}(p)$ the unitary representation of a general Lorentz transformation $\Lambda$ and associated Wigner matrix $W(\Lambda p)$ is expressed as~\cite{weinberg1995quantum}
\begin{equation}
    U(\Lambda) \, \Psi_{\sigma}(p)= N (p_0, \Lambda) \exp{(i \sigma \phi_W)} \Psi_{\sigma}(\Lambda p),
\end{equation}
where $N(p_0, \Lambda)$ is a normalization factor. The Wigner angles gives rise therefore to an overall phase for states with given helicity $\sigma$ but to a relative phase between the two components for states which are linear combinations of different helicity states. We also note that this phase is a function of the three rotation axes $\hat{n}_1, \hat{n}_2, \hat{n}_3$ defined by the sequence of rotations and not exclusively of the overall rotation matrix angle.  

\section{Deriving the axis and angles of the rotations}

The rotation $R_{\hat{z} \xrightarrow{} \hat{p}} (\hat{n_1}, \phi_1)$ taking $\hat{z}$ to $\hat{p}$ 
is not unique. Indeed, there is a family of possible rotations $R_{\hat{z} \xrightarrow{} \hat{p}} (\hat{n_1}, \phi_1)$, whose axis $\hat{n_1}$ is placed on the intersection between the unit sphere containing  $\hat{z}$ and $\hat{p}$ and the plane containing the bisector ($r$) and the vector  perpendicular ($s$) to  $\hat{z}$ and $\hat{p}$ as shown in Fig.~\ref{fig:rotFamily}. 
This can be shown by decomposing the rotation axis in a component in the plane $\hat{z}-\hat{p}$, a  component orthogonal to this plane and a null component perpendicular to both vectors. 

Indeed, we may solve Rodrigues' rotation formula for generic vectors $\hat{k}$ and $\hat{k'}$ as
\begin{equation}
    \hat{k'}  = \hat{k} \cos(\theta) + ( \hat{n} \, \cross \hat{k}  ) \sin (\theta) + ( \hat{n} \cdot \hat{k} ) (1- \cos(\theta) \hat{n} \label{rodriguesform}
\end{equation}
as an equation in the rotation axis $\hat{n}$. 
To this end, we express the generic rotation axis $\hat{n} $ as a linear combination of appropriate basis vectors 
\begin{eqnarray}
&& \hat{u}_1 = \frac{\hat{k}  \, \cross \hat{k'} }{|| \hat{k}  \, \cross \hat{k'} ||}   \\ \nonumber
&& \hat{u}_2 = \frac{ \hat{k}  + \hat{k'} }{||\hat{k}  + \hat{k'}  ||}  \\ \nonumber
&& \hat{u}_3 =  \frac{\hat{k'} - \hat{k} }{|| \hat{k'}  - \hat{k} ||}   .
\end{eqnarray}
The vectors $\hat{u}_1, \hat{u}_2$ are known solutions to (\ref{rodriguesform}), while the vector $\hat{u}_3$ completes the orthonormal basis formed by the first two. Upon decomposing the rotation axis along these three vectors 
\begin{equation}
    \hat{n} = \alpha \hat{u}_1 + \beta \hat{u}_2 + \gamma \hat{u}_3, 
\end{equation}
with $|\alpha|^2 + | \beta |^2 + |\gamma |^2 = 1$
it is easily shown that under non-trivial conditions for the rotation angle, the coefficient $\gamma$ is null, yielding $\hat{n}$ as a function of $\hat{u}_1, \hat{u}_2$. In our case we apply the above considerations to $\hat{z}, \hat{p}$ as initial and final vectors, respectively. This proves that any rotation with fixed initial and final vectors has null component in the  direction orthogonal to  the plane formed by $\hat{z} \cross \hat{p}$ and the bisector of $\hat{z}, \hat{p}$.

The family of rotation axes is thus described by the angle   
$\theta$ between $\hat{n_1}$ and the plane containing $\hat{z}$ and $\hat{p}$, i.e., 
\begin{equation}
    \hat{n_1}= r \cos(\theta) + s \sin(\theta). \label{eq:theta}
\end{equation}

A similar formula relating Wigner angle to the two boosts' directions is reported in~\cite{caffarelli}.

The angle $\theta$ is geometrically related to $\hat{n_1}$ to the rotation angle $\phi_1$ as:
\begin{equation}
    \sin(\theta) = \frac{\tan(\phi_{\hat{z}\xrightarrow{} \hat{p}}/2)}{\tan(\phi_1/2)}, 
\end{equation}
where $\phi_{\hat{z}\xrightarrow{} \hat{p}}= \acos(\hat{z} \cdot \hat{\hat{p}})$ is the angle between the $\hat{z}$ and $\hat{p}$, which is equivalent to $\phi_1$  when the rotation axis is perpendicular to $\hat{z}$ and $\hat{p}$.

Once the rotation $R_{\hat{z} \xrightarrow{} \hat{p}} (\hat{n_1}, \phi_1)$ is fixed, the inverse  rotation $R^{-1}_{\hat{z} \xrightarrow{}  \hat{\Lambda p}} (\hat{n}_3, \phi_3)$ is uniquely determined as it conserves the angle $\theta$, i.e., 
\begin{equation}
    \sin(\theta) = \frac{\tan(\phi_{\hat{z}\xrightarrow{} \hat{\Lambda p}}/2)}{\tan(\phi_3/2)}, 
\end{equation}
leading to the relation 
\begin{equation}
    \frac{\tan(\phi_1/2)}{\tan(\phi_3/2)} =\frac{\tan(\phi_{\hat{z}\xrightarrow{} \hat{p}}/2)}{\tan(\phi_{\hat{z}\xrightarrow{} \hat{\Lambda p}}/2)}.
\end{equation}

The rotation  $\Omega (\hat{n}_2, \phi_2)$ is generated by the product of two boosts, one in the direction $\hat{p}$ and the other in the direction $\hat{b}$, as shown in Eq.~(\ref{eq:polarDecomp}). Thus, the rotation axis of $\Omega$ is orthogonal to the directions of the two boosts, i.e., 
\begin{equation}
\hat{n_2} = \frac{\hat{p} \cross \hat{b}}{ \left\lVert \hat{p} \cross \hat{b}\right\rVert} .    \label{eq:n2}
\end{equation}
The rotation angle $\phi_2$ can be computed using the formula in~\cite{caffarelli}:
\begin{equation}
    \cos(\phi_2) = 1 - \frac{(\gamma_{\hat{b}}-1)(\gamma_{\hat{z}}-1)}
    {\gamma_{\hat{b}}\gamma_{\hat{z}} (1+ v_{\hat{b}} v_{\hat{z}} \hat{b} \cdot \hat{p})} 
    \left( 1-(\hat{b} \cdot \hat{p})^2 \right),
    \label{eq:phi2}
\end{equation}
where $v_{\hat{b}}$ and $ v_{\hat{z}}$ are the relative velocities related to boosts $\Lambda_{\hat{b}}$ and $B_{\hat{z}}$ (or equivalently  $B_{\hat{p}}$), respectively, while $\gamma_{\hat{b}}=1/ \sqrt{1-v_{\hat{b}}^2/c^2}$ and $\gamma_{\hat{z}}=1/ \sqrt{1-v_{\hat{z}}^2/c^2}$  are the respective Lorentz factors with $c$ being the light speed.

\section{The  case of orthogonal rotation axes}

Let us now consider the case in which the rotation axes are orthogonal to the vectors  $\hat{z}$, $\hat{p}$, and  $\hat{\Lambda p}$.
More specifically,  the three rotation axes $\hat{n_1}, \hat{n_2}, \hat{n_3}$ are now defined as the unit vectors parallel to $\hat{z}\times\hat{p}, \, \hat{p}\times\hat{\Lambda p}, \, \hat{\Lambda p}\times \hat{z}$, following the 'right-handed' convention of rotating the first vector onto the second one.  
Hence, the rotation axes and angles of $R_{\hat{z} \xrightarrow{} \hat{p}}$ and $R^{-1}_{\hat{z} \xrightarrow{}  \hat{\lambda p}} $  are respectively:  
\begin{eqnarray}
\hat{n_1} &= \frac{\hat{z} \cross \hat{p}}{ \left\lVert \hat{z} \cross \hat{p}\right\rVert}  &\quad \cos(\phi_1) = \hat{z} \cdot \hat{p}     \\
\hat{n_3} &= \frac{\hat{\Lambda p} \cross \hat{z}}{ \left\lVert \hat{\Lambda p} \cross \hat{z}\right\rVert} .    &\quad  \cos(\phi_3) = \hat{\Lambda  p} \cdot \hat{z}.    
\end{eqnarray}
For the rotation  $\Omega (\hat{n}_2, \phi_2)$, the axis $\hat{n_2}$ is always orthogonal to the vector $\hat{p}$ and $\hat{\Lambda p}$, as given in Eq.~(\ref{eq:n2}).

\begin{figure}[htb]
      \includegraphics[width=.4\textwidth]{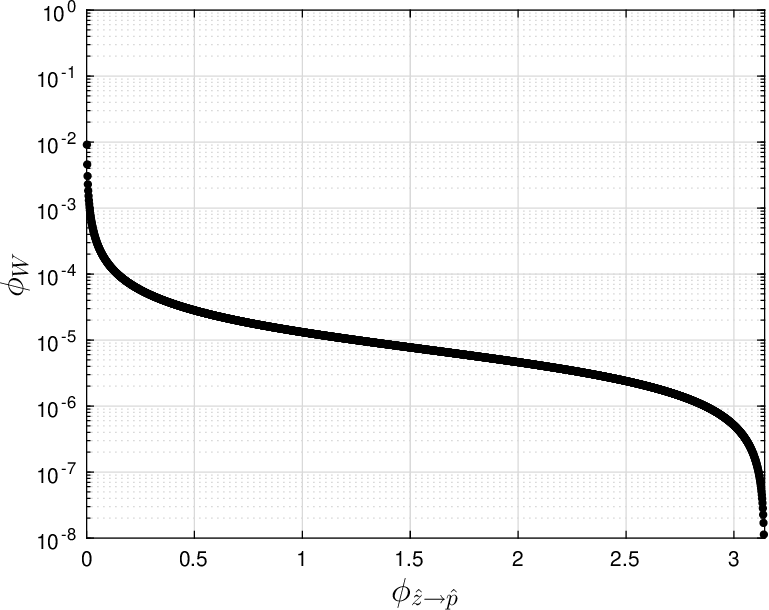}
    \caption{Wigner angle ($\phi_W$ in radian) as a function of the angle ($\phi_{\hat{z} \xrightarrow{} \hat{p}}$) between $\hat{z}$ and $\hat{p}=[\sin(\phi_{\hat{z} \xrightarrow{} \hat{p}}), 0, -\cos(\phi_{\hat{z} \xrightarrow{} \hat{p}})]$, for  $v_{\hat{z}} = 2/3 c$ and $v_{\hat{b}} = 8$ km/s. 
    }
    \label{fig1}

\end{figure}

In this case, the arcs corresponding to the rotation cosines lie along grand circles of the unit sphere. Therefore, the product of the three rotations defines a spherical triangle on the unit sphere (Fig.~\ref{fig1}), for which well known relations between the angles and the edges exist~\cite{todhunter}. The angles $\alpha_{ij}$ between the arcs $i$ and $j$ can be obtained from the scalar product of the rotation axes $n_i$ and $n_j$, i.e.,
\begin{equation}
    \pi - \alpha_{ij} = \acos(n_i \cdot n_j),
\end{equation}
taking into account that sides and angles of the polar triangle are, respectively, the supplements
of the angles and sides of the original triangle~\cite{todhunter}, Chap III. It is interesting to note, in fact, that by definition of polar triangle corresponding to a given spherical triangle, the rotation axes $\hat{n}_1,\hat{n}_2, \hat{n}_3$ identify the vertices of the polar triangle corresponding to the original points $\hat{z}, \hat{p}, \hat{\Lambda p}$.

The Wigner angle can be  derived in a straightforward manner from the spherical triangle, by resorting to the Gauss-Bonnet theorem~\cite{Yang-G-B,Hale-G-B}:
\begin{equation}
    \int_{B} K \, dA + \int_{\partial B} k ds = 2 \pi \chi (B),
\end{equation}
where $B$ is a bounded region on a surface in $\mathbb{R}^3$ and $\chi$ is the \textit{Euler characteristic} of $R$.
According to this theorem, when the unit sphere is rolled on a surface in such a way that the point of contact  moves along the arcs of the spherical triangle,
without slipping, then the result of this motion is a translation of the sphere together with a rotation
about an axis perpendicular to the surface and an angle equal to the excess of the spherical triangle~\cite{mukherjee2005simple}.

\begin{figure}[htb]
    \includegraphics[width=.4\textwidth]{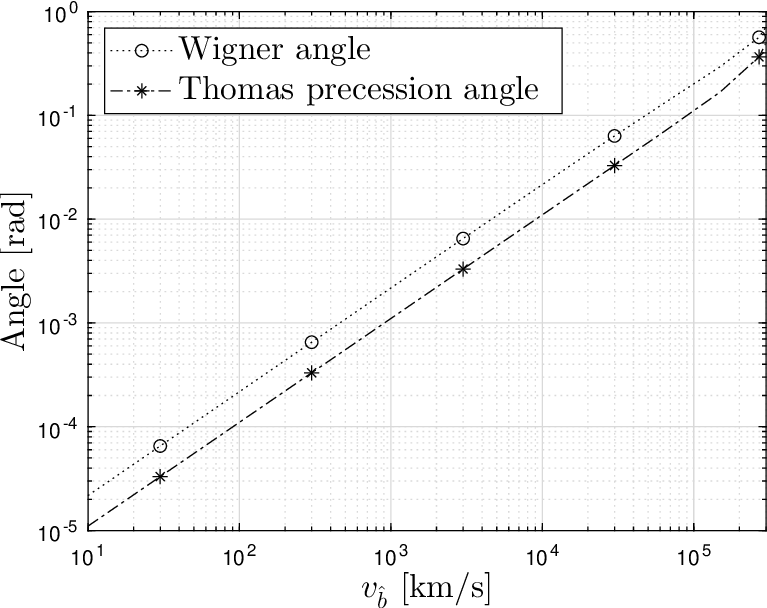}
    \caption{Wigner angle ($\phi_W$) for varying values of  $v_{\hat{b}}$, for  $\hat{p}  = [\sin(\pi/4), 0, -\cos(\pi/4)]$, $\hat{b} = [\cos(\pi/4),\sin(\pi/4),0]$ and   $v_{\hat{z}} = 2/3 c$.} 
    \label{fig2}
\end{figure}

\begin{figure}[htb]
    \includegraphics[width=.4\textwidth]{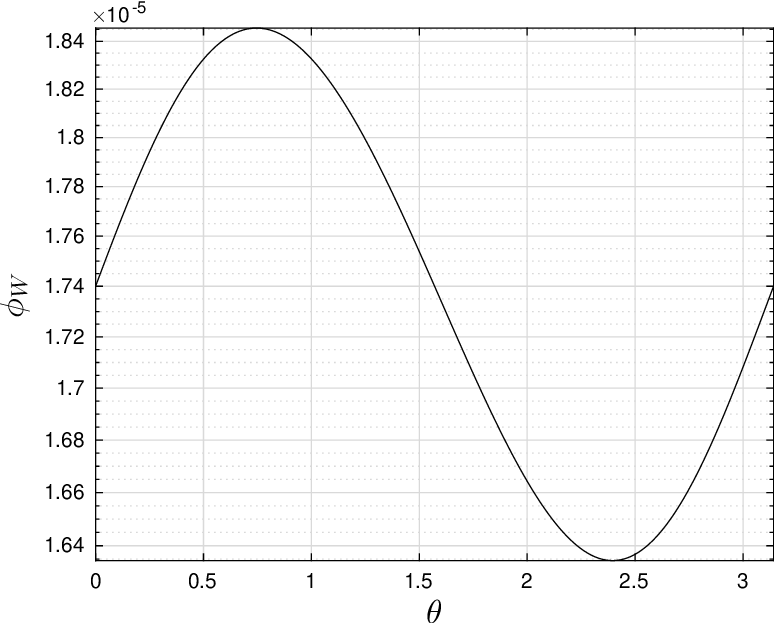}
    \caption{Wigner angle ($\phi_W$ in radian) for the family of rotations parametrized by the angle $\theta$, in radian, for $\hat{p} = [\sin(\pi/4), 0, -\cos(\pi/4)]$,  $\hat{b} = [\cos(\pi/4),\sin(\pi/4),0]$,   $v_{\hat{z}} = 2/3 c$ and $v_{\hat{b}} = 8$ km/s.} 
    \label{fig3}
\end{figure}

When applying the Gauss-Bonnet theorem by rolling the unit sphere from $\hat{z}$  to $\hat{z}$ along the spherical triangle, the Wigner rotation axis is thus $\hat{z}$ and the angle is the excess of the spherical triangle, i.e.,
\begin{eqnarray}
\hat{n_W}  &=&  \hat{z} \\
\phi_W &=&   [(\pi-\acos(\hat{n_1} \cdot \hat{n_2})) + (\pi-\acos( \hat{n_2} \cdot \hat{n_3})) \nonumber \\  
&+& (\pi-\acos ( \hat{n_1} \cdot \hat{n_3}))] -\pi.
\end{eqnarray}
Although this is a direct application of the Gauss-Bonnet theorem, an alternative proof of the excess angle formula is given as a result of three consecutive rotations forming a spherical triangle~\cite{mukherjee2005simple}. Alternatively, by using the `Lhuilier's formula'~\cite{lhuilier1809elemens}, the excess angle can be computed as a function of the edges of the spherical triangle as
\begin{eqnarray}
    \phi_W &=&  4 \arctan \left(\tan(\frac{s}{2}) \cdot \tan\left(\frac{s-\hat{z} \cdot \hat{p}}{2}\right)  \right.\\ 
        &\cdot&  \left. \tan \left(\frac{s-\hat{p} \cdot \hat{\Lambda p}}{2} \right) \cdot \tan \left(\frac{s-\hat{\Lambda p} \cdot \hat{z}}{2}\right) \right)^{1/2} \nonumber
\end{eqnarray}
where $s$ is the half-sum of the edges (or half-sum of the rotation angles), i.e., $s  = (\hat{z} \cdot \hat{p} + \hat{p} \cdot \hat{\Lambda p} + \hat{\Lambda p} \cdot \hat{z}) /2$.

\section{Numerical example}

The Wigner angle is assessed by considering the vectors $\hat{p}  = [\sin(\pi/4), 0, -\cos(\pi/4)]$  and $\hat{b}  = [\cos(\pi/4),\sin(\pi/4),0]$, unless otherwise stated. The relative velocities are assumed $v_{\hat{z}} = 2/3 c$ with $c$ the light speed and, unless otherwise stated,  $v_{\hat{b}} = 8$ km/s, which is equivalent to the typical satellite velocity of a low-Earth orbit (LEO) satellite in space.

Let us first consider the case of orthogonal rotation axes for all the  rotations. The position of $\hat{p}$ is parametrized by the angle $\phi_{\hat{z} \xrightarrow{} \hat{p}}$ (x-axis) as follows $\hat{p} = [\sin(\phi_{\hat{z} \xrightarrow{} \hat{p}}), 0, -\cos(\phi_{\hat{z} \xrightarrow{} \hat{p}})]$.
The Wigner angle 
changes as function of the position of $\hat{p}$ as displayed in Fig.~\ref{fig1}. The points are obtained analytically using the formula in Eq.~(\ref{eq:phi_W}). 

When varying the relative velocity $v_{\hat{b}}$, the Wigner angle decreases linearly with $v_{\hat{b}}$ (Fig.~\ref{fig2}). The figure shows also the Thomas precession angle obtained from Eq.~(\ref{eq:phi2}), which is one of the three contributions to the Wigner angle. We would like also to point out that the numerical values obtained for the Wigner angle are aligned with those derived by~\cite{noh2021quantum} from general relativity effects. We observe a similar behaviour for relativistic time corrections due to general and special relativity effects in Global Satellite Navigation Systems at the same altitude.

Fig.~\ref{fig3} displays the Wigner angle for the family of rotations (taking $\hat{z}$ to $\hat{p}$) parameterized by the angle $\theta$, as defined in Eq.~(\ref{eq:theta}).  The Wigner angle oscillates with $\theta$, although in a limited range. These variations indicate  that for a given $\hat{p}$ and boost $\Lambda$, the Wigner angle is not unique but depends on the rotation of the standard Lorentz transformation $L(p)$.

\section{Conclusions}
In this work we  derived an exact expression for the Wigner angle relating it to geometrical properties of spherical trigonometry. The sequence of three rotations defining the angle induces a geometrical phase on the quantum state of a massless particle.
This is a function of the sequence of rotation axes due to the overall Lorentz transformations. For given initial and final momenta of the particle, the Wigner angle is non-unique but rather a function of the path in momentum space, whose value can be geometrically computed by spherical geometry properties. 

Future research directions include  the extension of this work to entangled photon states 
~\cite{gingrich2003entangled} and  topological quantum materials ~\cite{kang2025measurements}.
Indeed, the importance of geometrical phases goes beyond the Lorentz transformation and the need to gain further insights on their physical, mathematical and conceptual nature.

\begin{acknowledgments}

This work was funded by the European Commission as part of the JRC work programme, Project 34039 Quantum-SGA. We are grateful to Adam Lewis for the insightful discussions and help.

\end{acknowledgments}
\bibliographystyle{apsrev4-2}
\bibliography{references}
\end{document}